\documentclass[preprint2]{aastex631}
\usepackage{longtable}
\usepackage{amsmath}
\usepackage{todonotes}
\usepackage{subfigure}
\usepackage{multirow}
\usepackage{appendix}
\usepackage{longtable}

\usepackage{hyperref}

\shorttitle{}
\shortauthors{Dornan \& Harris}

\graphicspath{{./}{figures/}}

\begin{document}

%\linenumbers

\title{Globular Cluster Systems in Dwarf Galaxies: Catalogs and Comparisons}

\author[0000-0002-7731-1291]{Veronika Dornan}
\affiliation{Institute for Astronomy, University of Edinburgh, Royal Observatory, Blackford Hill, Edinburgh, UK, EH9 3HJ}

\author[0000-0001-8762-5772]{William E. Harris}
\affiliation{Department of Physics and Astronomy, McMaster University, Hamilton, ON, L8S 4M1}

\begin{abstract}
The connection between a galaxy's total globular cluster system (GCS) mass and its halo mass has been studied for decades and it has been found that galaxies at nearly all observed masses adhere to a linear scaling relation between these properties. However, while we have ample, homogeneous data for galaxies with halo masses $M_h \gtrsim 10^{10} M_{\odot}$ the data available for low-mass galaxies is more sparse, and both GCS mass and halo mass estimates are determined using varying methodologies. This work compiles all available literature data for dwarf galaxies with confident stellar mass and GC count estimates, and converts these estimates to GCS masses and peak halo masses using a standard conversion. This allows for a consistent comparison of these masses to be made and a complete study of the behaviour of the $M_{GCS}-M_h$ relation to be conducted. We compare the positions of classical dwarfs on the scaling relation to that of ultra diffuse galaxies and extremely low-surface brightness galaxies and find that these non-classical dwarfs have, on average, systematically higher GC specific frequencies. This also makes them, on average, systematically positively offset from the $M_{GCS}-M_h$ relation, driving much of the high-$M_{GCS}$ scatter observed. 
\end{abstract}

\keywords{galaxies: star clusters -- galaxies: dwarfs -- galaxies: formation -- galaxies: evolution -- globular clusters: general}

\section{Introduction}\label{sec:intro}

Globular clusters (GCs) are a useful and frequently studied observational tracer of galaxy evolution. These dense stellar systems of up to $10^7$ stars are some of the oldest objects in their host galaxies, with typical ages between 10 and 13 Gyrs \citep{Vandenberg13, Beasley20}. Their compact sizes and high luminosities also allow them to be easily detectable at large distances \citep{Marta23, Joschko24}, which, when combined with their ubiquity \citep{Harris13, Ngoc25}, means the number of GCs hosted are often included in photometric analyses of galaxies.

The properties of a galaxy's GCS today can be closely linked to the evolutionary history the host galaxy as undergone. It has been shown that GCSs can grow through galaxy mergers \citep{Maji17,Lahen19,Newton24,Dornan25}, or can also be destroyed via mergers or tidal interactions with other, more massive galaxies \citep{Gnedin99, Kruijssen12a}. Comparing the global properties of galaxies to the properties of their GC systems (GCSs) can give clues about their merger histories or previous tidal interactions.

One way to probe the connection GCSs have with their host galaxies is through a galaxy's position on the GCS - halo mass ($M_{GCS}-M_h$) scaling relation. This linear relation between the mass of a galaxy's GCS and its total halo mass has been studied extensively, both through observations \citep{Blakeslee97, Blakeslee99, Spitler09, Hudson14, Harris17,Forbes18, Dornan25, Saifollahi25b} and simulations \citep{Kravtsov05,Kruijssen15,El-badry19,Choksi19,Bastian20, Valenzuela21}. This relation can tell us if certain classes of galaxies have experienced events that formed or accreted excess GCs for their masses \citep{Forbes25,Dornan25}, and what those events may have been. 

While massive galaxies are found to consistently (and tightly) follow the GCS mass - halo mass scaling relation \citep{Harris17,Dornan23,Dornan25}, the scatter observed in this relation for dwarf galaxies increases significantly \citep{Georgiev10,Forbes18, Prole19}. As a result, while the GCS mass for a galaxy with a stellar mass $\gtrsim 10^{10} M_{\odot}$ can be well predicted using this relation, dwarf galaxies have a wide range of possible GCS masses per unit stellar mass \citep{Burkert20,Eadie22}. Unlike massive galaxies, some dwarf galaxies can deviate very strongly from the expected $M_{GCS}-M_h$ relation by hosting no GCs at all \citep{Eadie22, Berek24}. While $N_{GC} = 0$ should not be unexpected
%this can be somewhat expected 
for very low-mass dwarfs with stellar masses below $M_{\star} \lesssim 10^{7} M_{\odot}$, where the expected average $N_{GC}$ is smaller than the dispersion in the relation, puzzlingly galaxies with stellar masses as high as $M_{\star} \sim 10^9 M_{\odot}$ have been found to lack GCs \citep{Peng08,Georgiev10}. It is currently unclear if these galaxies formed without GCs, once hosted GCs which have now been destroyed, or host very faint or obscured GCs that have yet to be detected.

In order to fully understand the cause of this wide range in GCS masses for dwarfs, it is necessary to isolate the scatter in the $M_{GCS}-M_h$ relation that is caused by physical processes affecting dwarf galaxies and their GCSs from the scatter caused by differing observational techniques between surveys and studies. While there are many excellent studies of dwarf galaxy GCSs (see Section \ref{sec:surveys}), it can be difficult to compare the data from these studies to one another on the $M_{GCS}-M_h$ relation as the methods used to determine total number of GC ($N_{GC}$) estimates, GCS masses, and halo masses can vary. This work aims to compile the data from surveys and catalogs already available in the literature and clearly lay out the differing methods applied in the photometric analyses of the GCSs and determine $M_{GCS}$ and $M_{h}$ estimates for these galaxies using consistent methods. This not only allows for the positions of these galaxies on the $M_{GCS}-M_h$ relation to be properly compared to one another, but also provides a convenient, public catalog of the data from several major dwarf galaxy GCS surveys.

In this work we list the surveys and catalogs included in our combined catalog and compare the methods employed by each to obtain their $N_{GC}$ estimates in Section \ref{sec:surveys}. We then detail the scaling relations used to determine GCS and peak halo masses for all galaxies in our combined catalog in Section \ref{sec:masses}. With this data, we then plot the positions of all galaxies in the combined catalog on the $M_{GCS}-M_h$ relation as well as the $N_{GC}-M_{\star}$ and $N_{GC}-M_h$ scaling relations as well, and compare the positions of classical dwarfs to those for ultra diffuse galaxies (UDGs), extremely low surface-brightness galaxies (ELSBGs, see Section \ref{sec:UDGs/ELSBGs}), and massive galaxies in Section \ref{sec:results}. Finally we discuss the implications of these results and summarize our findings in Sections \ref{sec:discussion} and \ref{sec:summary} and briefly discuss necessary future work in this field in Section \ref{sec:future}.

\section{Catalog and Individual Survey Notes} \label{sec:surveys}

This literature catalog is comprised of seven dwarf galaxy surveys and previously published literature catalogs. These catalogs vary in the data available for their GCSs and host galaxies, but all have $N_{GC}$ estimates as well as total galactic V-band absolute magnitudes and/or total stellar masses. For the sake of purity of the combined sample, some data from the original surveys and catalogs have been omitted from the analysis done in this work. Data with the following criteria have been culled:
\begin{enumerate}
    \item \textbf{Systems with $N_{GC}$ estimates $\leq 0$:} While there exist a large number of dwarf galaxies which host no GCs (see Section \ref{sec:intro}), in this work we will be considering only GC-hosting galaxies. This is because it is still unclear whether or not current non-GC-hosting galaxies originally hosted GCSs and have since had them fully disrupted, and/or because current non-GC-hosting galaxies may in fact host some low-mass, dim GCs that we are currently unable to reliably detect. As such, at this time we are omitting these galaxies from the combined catalog.
    \item \textbf{Systems with $N_{GC}$ uncertainties greater than $N_{GC}$ estimates:} In line with the above reasoning, we are restricting our combined sample to galaxies with $N_{GC}$ estimates with at least \textit{one} confident GC candidate.
\end{enumerate}

The above criteria will bias our GCS catalog to somewhat more massive dwarf GCSs. Future steps to address this bias and include systems with no GCs are discussed in Section \ref{sec:future}.

Below we lay out the differing photometric methods and assumptions used by each survey and catalog included in our combined catalog. At the end of this section we summarize the important differences and similarities that should be taken into consideration when comparing the results of each of these GCS samples.

\subsection{ACS Virgo Cluster Survey and ACS Fornax Cluster Survey}

The data from the ACS Virgo Cluster Survey are taken from \cite{Peng08}, which studied the GCSs of early-type galaxies in the Virgo cluster, resulting in a large, homogeneous GCS catalog. This catalog spans a large range of host galaxy stellar masses, so we only include 63 galaxies with $M_{\star} \leq 2\times 10^{10} M_{\odot}$ in our combined catalog. The data from the ACS Fornax Cluster survey are taken from \cite{Liu19}, which used nearly identical methods and imaging to the ACS Virgo Cluster Survey but applied to the Fornax Cluster, functionally homogeneously extending the Virgo survey. We include 16 galaxies from this survey in our combined catalog.

All imaging for these two surveys was conducted with the Hubble Space Telescope (HST) advanced camera for surveys (ACS) using the F475W and F850LP filters, which roughly correspond to the Sloan Digital Sky Survey (SDSS) \textit{g} and \textit{z} bands. GC candidates were only included if they were detected in both filters. GC candidates were selected on the basis of colour, size, and magnitude. They selected objects with colours between $ 0.5 < g-z < 2.0$, which is a wide range intended to include the full range of ages and metallicities typical of old star clusters. They next plotted all potential GC candidates on a size-magnitude diagram and determined a maximum likelihood estimation for each galaxy's objects to assign a GC probability value. Only objects with a GC probability $> 0.5$ were selected.

Finally, to obtain a $N_{GC}$ estimate they conducted completeness corrections based on their limiting magnitudes and adopted dwarf galaxy GC luminosity function (GCLF). This correction was minimal, as they estimate that the deepness of their imaging recovered the brightest $\sim 90\%$ of each galaxy's GCSs. They adopted the GCLF derived in \cite{Jordan07}, which was determined through stacking GC photometry data from 89 of the galaxies in the ACS Virgo Cluster survey. 

The stellar masses of these galaxies were determined using $M_{\star}/L_B$ mass-to-light ratios determined for each galaxy from the \cite{Bruzual03} models. To determine these ratios they used a combination of their $g-z$ colours and \textit{J - K} colours from 2MASS (the Two Micron All Sky Survey Extended Source Catalog).

\subsection{Fornax Deep Survey}\label{subsec:deep_fornax}

The data from the Deep Fornax Cluster Survey are taken from \cite{Prole19}, which studied the GCSs of low surface brightness galaxies around the central brightest cluster galaxy of the ACS Fornax Survey. We include 170 galaxies from this survey in our combined catalog.

The imaging for these galaxies was obtained using the VLT Survey Telescope/OmegaCAM instrument in the \textit{u, g, r} \& \textit{i} bands. GC candidates were selected based on ellipticity, magnitude, and colour. Objects were selected if they had minor-to-major axis ratios greater than $(b/a)> 0.95$, \textit{g-}band apparent magnitudes greater than $m_g > 19$, and had colours within the following ranges: $-0.18 < g-r < 1.23$, $0.32 < g-i < 2.00$, and $0.37 < u-g < 5.07$.

Total GC distributions were modelled using a  Bayesian mixture model, represented by a Plummer profile, to represent the GCs associated with the target galaxy, plus a uniform distribution, to represent background GCs. This model had two free parameters: the half-GC radius of the galaxy, and the fraction of all the GCs which belong to the target galaxy. The MCMC code EMCEE \citep{emcee} was run on this model to determine the most likely values for these parameters, which were then used to calculate the total number of GCs hosted by each galaxy.

They also applied GC completeness corrections adopting a GCLF with the same form as in \cite{Villegas10} which studied the GCSs of Fornax galaxy dwarfs, although at brighter total magnitudes than the dwarfs in the Fornax Deep Survey. They estimate their GC completeness to be between $\sim60\% - 90\%$, resulting in somewhat higher corrections needed for their lower-luminosity galaxies than was needed for the other surveys. The stellar masses of their galaxies were determined using the $M_i$ magnitudes and $g-i$ colours, converted using equation 8 in \cite{Talor11}, re-stated below in equation \ref{eq:taylor}.

\begin{equation}\label{eq:taylor}
    \log(M_{\star}/[M_{\odot}]) = 1.15 + 0.70(g-i) - 0.4 M_i
\end{equation}

\subsection{ELVES Survey}

The data from the Exploration of Local VolumE Satellites (ELVES) Survey are taken from \cite{Carlsten22}, which studied the GCSs of early-type satellites of Milky Way-like and small group hosts in the Local Volume. We include 25 galaxies from this survey in our combined catalog.

The imaging for these galaxies came from a mixture of archival CFHT/MegaCam, DECam Legacy Survey (DeCaLs), and Subaru/Hyper Suprime-Cam imaging. They note that the CFHT and Subaru imaging are both deeper than the DeCaLs imaging. All galaxies were imaged in the \textit{g-}band as well as either the \textit{i-}band or \textit{r-}band. The individual filters used are all SDSS-like, but do differ somewhat between cameras. However the authors show that this difference is less than 0.1 magnitude and does not significantly affect their results. 

GC candidates were selected via colour and magnitude cuts. They restricted their candidates to objects with colours between $0.1 < g-r < 0.9$ or $0.2 < g-i < 1.1$, intended to cover the full range of dwarf galaxy GC colours observed in \cite{Prole19} (see section \ref{subsec:deep_fornax}). Candidates were also restricted to absolute g-band magnitudes between $-9.5 < M_g < -5.5$ to cover the expected GCLF. They then applied two different methods to determine GC abundance; a simple background selection and a likelihood-based inference. 

They perform completeness corrections assuming a \cite{Jordan07} GCLF for their galaxies, as they found that the properties of this GCLF are very similar to those for Local Volume and Virgo cluster dwarf galaxies. Their imaging is also deep enough that these corrections are minimal, resulting in $N_{GC}$ corrections of $\lesssim 5\%$. The stellar masses of their galaxies were determined using a mass-to-light ratio of $M_{\star}/L_g = 1.24$, derived from \cite{Into13} using $g-i =0.74$, the average galaxy colour in their sample.

\subsection{MATLAS Survey}

The data from the Mass Assembly of early-Type GaLAxies with their fine  Structures (MATLAS) Survey are taken from \cite{Marleau24}, which studied 74 UDGs within the larger survey. We include 27 galaxies from this study in our combined catalog based on our $N_{GC}$ estimate selection criteria.

The GC imaging for these galaxies was taken with HST using the ACS in the F606W and F814W filters, which roughly correspond to the SDSS \textit{r} and \textit{i} bands. We also used global colour information for these galaxies taken from \cite{Poulain21}, which imaged them using CFHT in the \textit{g} and \textit{r} bands (with the exception of MATLAS-342, which was imaged in the \textit{g} and \textit{i} bands).

GC candidates were selected based on colour and concentration. GCs were selected with colours between $0.5 < (m_{F606W} - m_{F814W})_0 < 1.2$ and concentration indices between $0.1 < \Delta m_{4-8}<0.5$ for galaxies within 25 Mpc and between $-0.1 < \Delta m_{4-8}<0.5$ for galaxies beyond 25 Mpc. The concentration indexes here represent the difference in apparent magnitude in the F606W band for a GC when observed using a 4 vs 8 pixel diameter aperture, corresponding to 0.1 and 0.2 arcseconds respectively. GCs were considered hosted by the target galaxy if they are within $2R_e$, and those found beyond $2R_e$ were used to determine GC background contamination. 

Finally, they performed a completeness correction assuming a GCLF of the form used in \cite{Miller07}, which was determined using dwarf galaxies in the Virgo and Fornax Clusters and the Leo Group. The stellar masses of these galaxies were not listed in \cite{Marleau24}, so they were determined independently here using the colours from \cite{Poulain21} and the colour-mass-to-light ratio relations in \cite{Into13}, for consistency with the ELVES survey stellar masses.

\subsection{Georgiev Catalog}

\cite{Georgiev10} compile the GCSs of 41 faint, late-type dwarf galaxies in low-density environment, of which we include 40 in our combined catalog. The only galaxy to be omitted from this catalog was the SMC, as since the publication of \cite{Georgiev10}, it has been determined that the SMC could actually be two distinct structures which are superimposed on each other in our line or sight \citep{Murray19,Murray24}. As a result, the stellar mass estimate of the SMC quoted in \cite{Georgiev10} is very likely an overestimate and should not be included in this combined dwarf catalog.

This catalog uses HST archival imaging in the F606W and F814W filters. GC candidates were selected based on colour, ellipticity, size, and concentration cuts. GCs were selected with colours between $-0.4 <
(m_{F606W} - m_{F814W})_{o} < 0.15$, ellipticities below $e < 0.15$, full width half maxima (FWHM) between $2 < FWHM < 9$ pixels, and concentration indexes above $\Delta m_{2-3} > 0.4$. 

They estimate, due to the depth of their imaging, that they have $\sim 90 \%$ completeness for the GCSs and as a result choose not to apply a GCLF-based completeness correction. They determined their galaxy stellar masses using the colour-mass-to-light ratio relations in \citep{Bell03}.

\subsection{Gannon Catalog}

\cite{Gannon24} compile the GCSs of 33 UDGs, of which we include 19 in our combined catalog based on our $N_{GC}$ estimate selection criteria. This is a literature catalog which drew from studies of individual galaxies, often times combining data from multiple studies of the same galaxy, and as such does not have any singularly standardized photometric parameters or GC detection methodology. As such, while the GC detection methods for these galaxies may vary, the $N_{GC}$ estimates can be taken as reliable lower bounds. In addition, half of the UDGs included here from the \cite{Gannon24} catalog had data taken from the same two studies ($N_{GC}$ values taken from \cite{Lim2020} and stellar masses taken from \cite{Toloba2023}), which would make them internally self-consistent. 

A full list of the original papers this \cite{Gannon24} catalog draws from can be found in Appendix \ref{app:Gannon_papers}. For more information on this data, please refer to the original papers.

\begin{center}
\begin{table*}[h!tb]
    \scriptsize
    \centering
    \caption{Survey Comparison} \label{tab:compare}
    %\hspace{-1cm}
    \begin{tabular}{|c|c|c|c|c|c|c|c|}
    \hline \hline
    Survey & Instrument & Mag Limit & Resolution & Distance & Filters & Colour Cuts & GCLF Peak \\
     &  &  & (``/pixel) & (Mpc) &  &  & \\
    \hline \hline
    ACS Virgo/Fornax & ACS & $m_g \sim 26$ & 0.05 & $\sim 16/19$  & \textit{g,z} & $0.5 < g-z < 2.0$  & $M_g = -7.2$ \\
    \hline
     &  &  &  &  &  & $-0.18 < g-r < 1.23$ & \\
    Fornax Deep & OmegaCAM & $m_g \sim 25$ & 0.21 & $\sim 19$ & \textit{u,g,r,i} &  $0.32 < g-i < 2.00$ &  $M_V = -7.4$ \\ 
     &  &  &  &  &  & $0.37 < u-g < 5.07$ &  \\
    \hline
     & HSC & & 0.4 & &  & $0.1 < g-r <0.9$&  \\
    ELVES & MegaCam & $m_g\sim 24$ & 0.19 & $< 12$ & \textit{g,r,i} & $0.2 < g-i <1.1$  &  $M_g = -7.2$ \\
     & DECam & & 0.26 &  &  &  &   \\
     \hline
    MATLAS & ACS & $m_g \sim 24$ & 0.05 & $17 -46$  & \textit{r,i} & $0.5 < r-i < 1.2$ &  $M_V = -7.3$ \\
    \hline
    Georgiev+ (2010) & ACS & $m_V \sim 26$  & 0.05 & $< 9$  & \textit{r,i} & $-0.4<r-i <0.15$ & N/A \\
    \hline \hline
    \end{tabular}
\item{} \footnotesize{\textit{Key to columns:} (1) Survey name; (2) Instrument used for imaging. ACS corresponds to HST, OmegaCam corresponds to VLT, HSC corresponds to Subaru, MegaCam corresponds to CFHT, and DECam corresponds to Víctor M. Blanco 4-meter Telescope; (3) Limiting GC magnitude of photometry; (4) Spatial resolution of photometry; (5) Distance to target galaxies; (6) SDSS-like filters used; (7) colour cuts applied for GC selection; (8) peak magnitude of the GCLF used for completeness corrections.}
\end{table*}
\end{center}

\subsection{Comparison of Surveys and Catalogs}

Broadly, the assumptions made for determining the GCS properties of the dwarf galaxies in each survey are similar enough to be compared to one another, however there are differences that can have systematic impacts on the final $N_{GC}$ estimates.
\begin{itemize}
    \item \textbf{Deepness of Imaging:} Deeper imaging allows for a more complete photometric census of the GCS. With deeper imaging, fainter GCs can be detected and completeness corrections become more minor.
    \item \textbf{Resolution of Imaging:} Higher resolution imaging also allows for smaller on-sky GC sizes to be reliably detected. Lower resolution imaging could result in smaller (and therefore lower mass) GCs to be less easily detected, skewing average GC properties to higher masses. Lower resolutions could also contribute to misidentifications of GC candidates, increasing the field contamination.
    \item \textbf{Distance to Target:} Related to the above, targets that are closer will not require as deep imaging or high resolution to detect GCs of the same faintness or size at further distances. 
    \item \textbf{Filters Used:} All of the surveys included in this study select GCs based on colour cuts, so the same GC could be included or excluded when using different filters. 
    \item \textbf{GC Candidate Selection:} In addition to differing filters, the specific criteria for GC selection can also vary, including the bounds for those criteria. Even with standardized GC criteria, however, there are also differing definitions for GC association with its target galaxy and methods of subtraction of background GCs.
    \item \textbf{GCLF Used for Completeness Corrections:} Finally, adopting a different GCLF will change the completeness correction applied to the $N_{GC}$ estimate. Thankfully, the majority of the surveys used here make minor corrections for completeness and used very similar GCLFs. 
\end{itemize}

A direct comparison of the photometry, GC selection criteria, target properties, and GCLFs used by each survey is listed in Table \ref{tab:compare}. Overall, differences between the limiting magnitudes and GCLFs used for completeness corrections for these surveys are very small, however, the surveys begin to differ in terms of resolution, distance, and colour cuts.

Three of the surveys use HST/ACS imaging, resulting in identical resolutions. The Fornax Deep and ELVES surveys use a variety of ground-based telescopes with similar resolutions, but which are all nearly 4 times lower than HST/ACS. In addition, there is little agreement between the surveys in term of the filters used for colour cuts. Those that do use the same filters still define their bounds quite differently, typically because they are trying to identify an area in a colour-colour or colour-concentration space for their GC selection. Each of these surveys make different trade-offs in terms of purity of their samples vs. completeness and should be considered when comparing the results in Section \ref{sec:results}.

\section{Mass Conversions} \label{sec:masses}

We plot given $N_{GC}$ and $M_{\star}$ estimates for all of these surveys and catalogs as is in Figure \ref{fig:star-GC}. Many works which have studied GCS scaling relations for dwarf galaxies focus on the $N_{GC}-M_{\star}$ relation as a way to avoid having to make mass conversions with large uncertainties to obtain the more well-studied $M_{GCS}-M_h$ relation. However, to properly connect and compare the GCSs of dwarf galaxies to their more massive counterparts, the more linear $N_{GC}-M_h$ or $M_{GCS}-M_h$ relation is needed. First, let us begin with converting stellar masses to halo masses.

\subsection{Stellar-to-Halo Masses}

As was mentioned in the previous section, different surveys and catalogs in this work have used slightly different mass-to-light ratios to obtain their galaxy stellar masses, however from now on in our analysis we will treat them as consistent. Obtaining accurate halo masses for dwarf galaxies can be much more difficult than for higher mass galaxies. Dwarf galaxies' smaller sizes result in less material in the outer regions to be used to get accurate velocity dispersions to estimate the total dynamical masses. Although some of the works included in this combined catalog determine the dynamical masses of their galaxies, they take wildly different approaches.

The galaxies included in the sample used in \cite{Forbes18} had stellar and HI gas kinematic data available to allow for mass estimates to be made. \cite{Prole19}, on the other hand, determined the halo masses of their sample using the very $N_{GC}-M_h$ relation we are studying in this paper. As it currently stands, the most accurate way to estimate total masses for dwarf galaxies requires deep spectroscopic imaging to obtain the most complete kinematic data possible \citep{Buzzo25b, Haacke25}, an approach which is difficult to apply to a large sample of galaxies, and which may still fail for the faintest dwarfs with low stellar membership. It should also be noted that even with high accuracy velocity dispersion measurements, extrapolation to large radii and adoption of an assumed DM model is still needed to estimate halo masses.

In order to be able to most accurately compare dwarf galaxies to one another on the $M_{GCS}-M_h$ relation, we will apply a simple, but standard, conversion from stellar mass to peak halo mass. While there have been many recent studies of the behaviour of the stellar-to-halo-mass-relation (SHMR) for dwarf galaxies \citep{Read17, Nadler20, Munshi21, Manwadkar22, Christensen24}, here we apply the SHMR modeled by \cite{Daneli23}. We choose to use this SHMR as it was determined using a semi-analytic model sampled from the ELVES survey, which is included in this combined catalog. This SHMR takes the following form:

\begin{equation}
    \begin{split}
    \log(M_{\star}) = 10.457 - \log(10^{-2.10 x} + 10^{-0.464x}) \\-0.812 \exp\Big[-0.5\Big(\frac{x}{0.319}\Big)^2\Big] 
    \end{split}
\end{equation}

where:

\begin{equation}
    x = \log\Big(\frac{M_{peak}}{10^{11.889}} \Big)
\end{equation}

For galaxies in our sample, this $M_{peak}-M_{\star}$ relation behaves linearly, with a constant scatter of $\sigma = 0.06^{+0.07}_{-0.05}$ \citep{Daneli23}. We incorporate this scatter into the error propagation for our peak halo mass estimates alongside the quoted uncertainties for the galaxies' stellar masses.

It should be acknowledged, however, that this method of halo mass estimation is dependent on the assumption that all galaxies in this catalog adhere to the SHMR. Some work which has determined the total halo masses of UDGs kinematically have found that this class of galaxy may have higher dynamical masses per unit stellar mass than classical dwarfs \citep{Forbes24, Forbes2025b}. If this is the case for a majority of UDGs or ESLBGs, it would affect their true positions on the $M_{GCS}-M_h$ relation, which is discussed further in Section \ref{sec:udgs}

We would also like to note that this SHMR is not necessarily between the stellar mass of the dwarf galaxy and its \textit{present-day} halo mass, but rather with its \textit{peak} halo mass. Other studies of the SHMR for dwarfs have found that when linking it to present-day halo mass there is significant degeneracy and scatter dependent on galaxies' interaction history with other, more massive galaxies \citep{Munshi21,Christensen24}. This degeneracy is significantly limited when instead comparing the current stellar mass to the highest halo mass the galaxy had in its lifetime (peak halo). Further discussion of the implications of this use of peak halo mass over current halo mass for the $M_{GCS}-M_h$ relation can be found in Section \ref{sec:discussion}. Regardless, we will still refer to the scaling relation as the $M_{GCS}-M_h$ relation in this work.

\subsection{$N_{GC}$-to-GCS Masses}\label{sec:GC_masses}

We also apply a simple, standard conversion from $N_{GC}$ to GCS mass using the properties of the GC luminosity function (GCLF), and by extension GC mass function (GCMF), observed in dwarf galaxies. We take the peak of the GCLF to determine an average GC mass for a given galaxy ($\langle M_{GC} \rangle$), with the uncertainty on that mass equivalent to $1\sigma$ on the GCLF. We then simply multiply our $N_{GC}$ estimates by $\langle M_{GC} \rangle$ to get an estimate of the total mass of the system. 

Other studies have taken a single GCLF peak and $\langle M_{GC} \rangle$ and applied it to all of their galaxies to obtain $M_{GCS}$ estimates \citep{Forbes18}. However, there exists a shallow relation between $\langle M_{GC} \rangle$ and host galaxy dynamical mass \citep{Harris13}, driven by the fact that the GCLF peak shifts to dimmer magnitudes for smaller galaxies \citep{Jordan07, Villegas10}. We adopt $\langle M_{GC} \rangle$ values for our galaxies based on this relation, as although the shift to lower average GC masses is small, for galaxies with only a handful of GCs it can more noticeably impact their position on the $M_{GCS}-M_h$ relation. This scaling relation takes the following form \citep{Harris13}:

\begin{equation}
    \langle M_{GC} \rangle = (2.26 \times 10^4) \times M_{dyn} ^{0.098}
\end{equation}

We quote the uncertainty in our $\log\langle M_{GC} \rangle$ estimate as the vertical dispersion in the $\langle M_{GC} \rangle - M_{dyn}$ relation, stated in \cite{Harris13} to be $\sigma_{\log \langle M_{GC} \rangle} = 0.086$. We incorporate this uncertainty into our error propagation for $M_{GCS}$ alongside the quoted $N_{GC}$ errors from the surveys and catalogs. For the \cite{Forbes18} and \cite{Georgiev10} catalogs no errors on their $N_{GC}$ estimates are stated, so we adopt $30 \%$ uncertainties.

An important fact to note about the use of this $\langle M_{GC} \rangle - M_{dyn}$ relation, is that here we use the peak halo masses that were obtained from the SHMR as an estimate of $M_{dyn}$. While the shallow dependence on $M_{dyn}$ of this relation means that the changes in $M_h$ from peak to present day results in only a minor change to the $\langle M_{GC} \rangle$ used, it may still not be entirely accurate.

\section{Results} \label{sec:results}

The original $N_{GC}$ and stellar mass estimates from the literature and the systemically calculated GCS and peak halos masses from this work are available in the complete catalog. This data can be accessed here: {\href{https://github.com/dornanv/Dwarf-Galaxy-GCS-Catalog}{\texttt{github.com/dornanv}}}

\subsection{Scaling Relations}

\begin{figure*}[h!tb]
    \centering
    \includegraphics[width=1\textwidth]{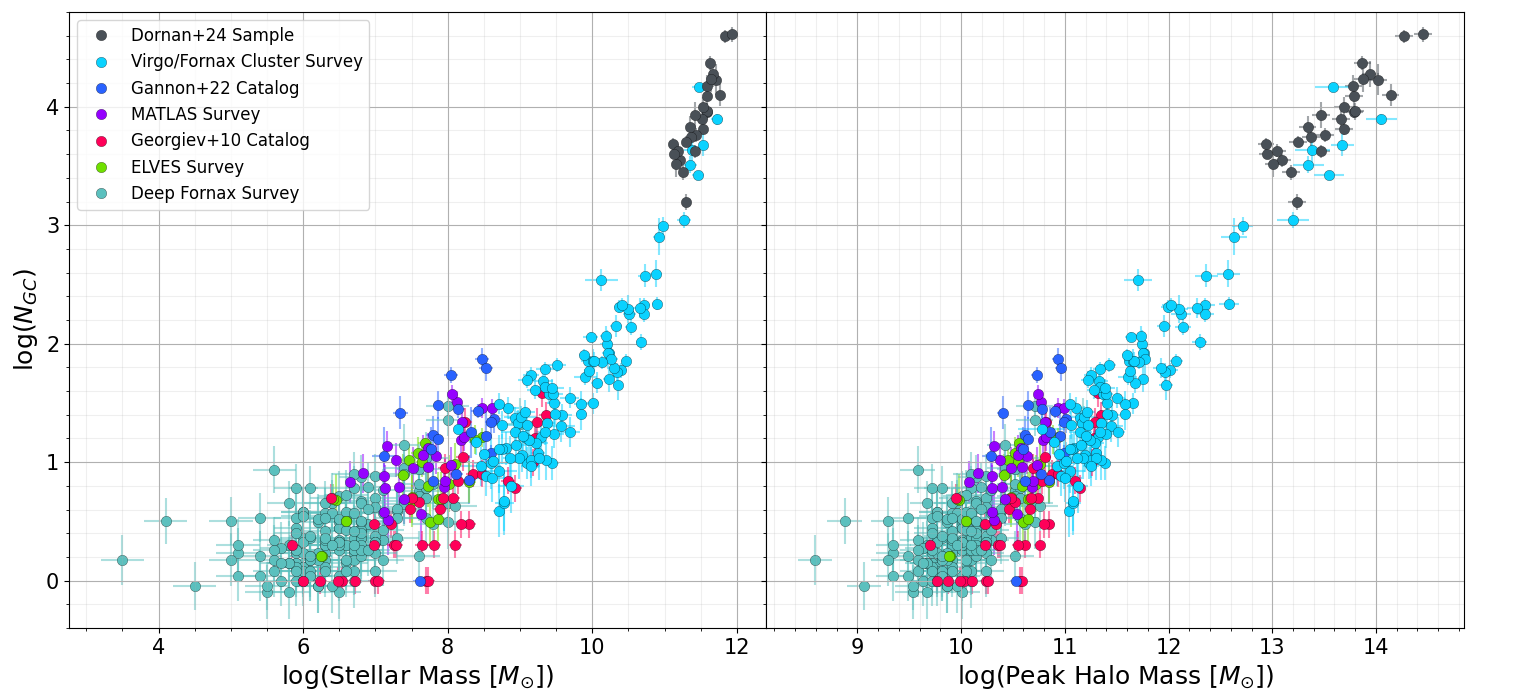}
    \caption{\label{fig:star-GC} \footnotesize{\textit{Left:} $N_{GC}$ vs host galaxy total stellar mass for all galaxies in the combined sample and the more massive galaxies in the Virgo Cluster Catalog \citep{Peng08} and the BCGs in the \cite{Dornan25} sample. \textit{Right:} Same as the left panel, but with stellar masses now converted to peak halo masses following the method outlines in Section \ref{sec:masses}.}}
\end{figure*}

Here we plot where the galaxies included in this combined catalog sit on three different scaling relations: $N_{GC}-M_{\star}$, $N_{GC}-M_h$, and $M_{GCS}-M_h$. The first, $N_{GC}$ vs stellar mass is commonly investigated by dwarf galaxy GCS studies \citep{Eadie22,Berek24}, and can be seen in Figure \ref{fig:star-GC}.  

While this scaling relation limits the assumptions necessary when converting $N_{GC}$ to $M_{GCS}$ and $M_{\star}$ to $M_h$, it is a globally non-linear relation, with a break-point at $M_{\star} \sim 10^{11}M_{\odot}$. This is due to the fact that although this scaling relation is a by-product of that between $N_{GC}$ and $M_h$, which remains linear even to the highest mass galaxies, this break in linearity arises from the same break in the non-linear SHMR. 

As such, it is more relevant to plot the $N_{GC}-M_h$ relation, as can be seen in Figure \ref{fig:star-GC}. Here, the scaling relation tightens and becomes linear across decades of peak halo mass. However, this linearity is still not fully universal. As we move to the lowest-mass dwarf galaxies in our combined catalog it is impossible to have fewer than $\sim1$ GC, resulting in a flattening of the relation. However, as was discussed in section \ref{sec:GC_masses}, the average mass of a GC in a galaxy scales with that host galaxy's dynamical mass.

\begin{figure*}[h!tb]
    \centering
    \includegraphics[width=1\textwidth]{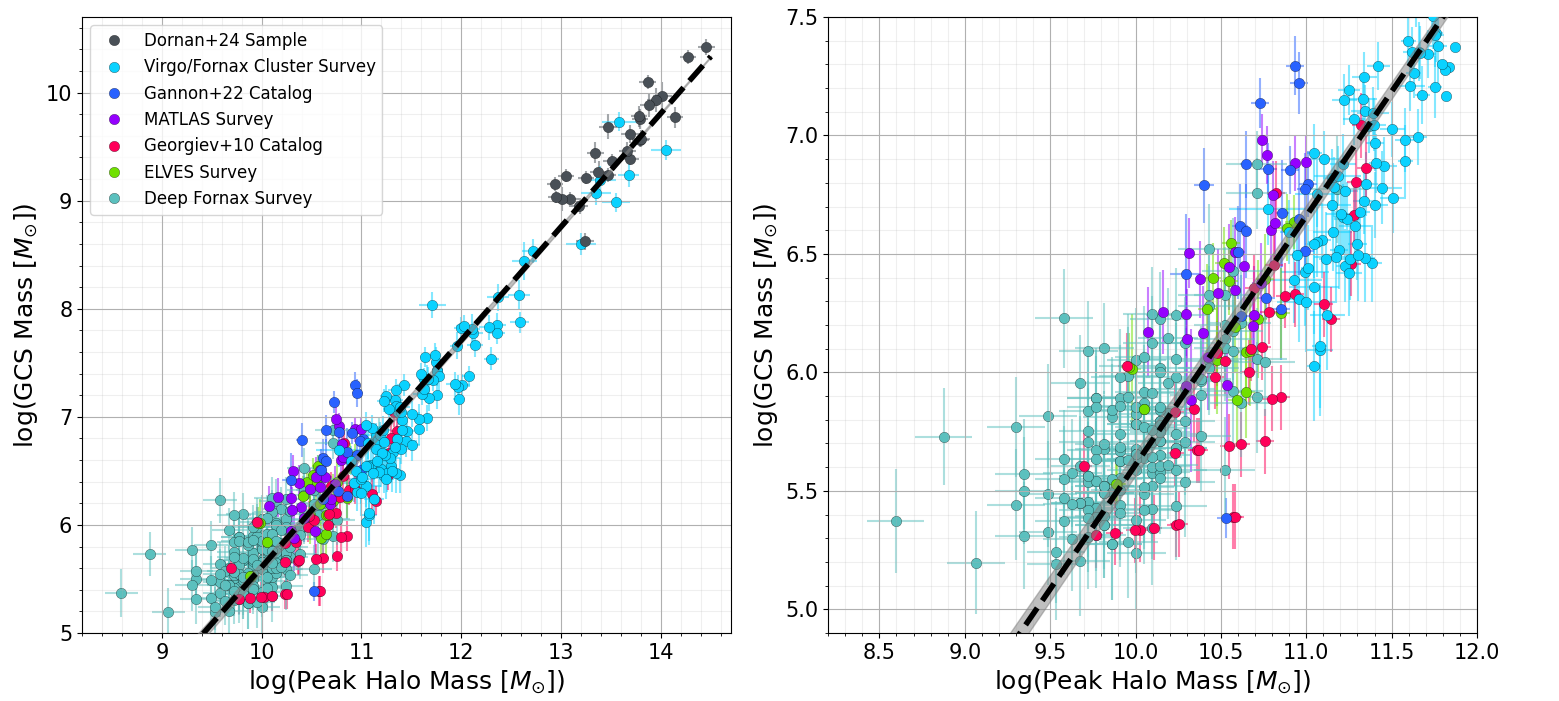}
    \caption{\label{fig:h-MGC} \footnotesize{\textit{Left:} $M_{GCS}$ vs peak $M_{h}$ for all galaxies in the combined sample and the more massive galaxies in the Virgo Cluster Catalog \citep{Peng08} and the BCGs in the \cite{Dornan25} sample. \textit{Right:} Same as the left panel, but zoomed-in on the combined dwarf galaxy catalog only.}}
\end{figure*}

In Figure \ref{fig:h-MGC} we plot the $M_{GCS}-M_h$ relation for our combined sample alongside the full ACS Virgo Cluster Survey and the \cite{Dornan25} BCG sample. The stellar masses adopted for these more massive galaxies were determined using the \cite{Hudson15} SHMR. This is because the \cite{Daneli23} SHMR was sampled from galaxies in the ELVES Survey, but did not include any galaxies with stellar masses above $M_{\star} > 10^{11} M_{\odot}$, with the majority of the galaxies having stellar masses well below that. However, the \cite{Hudson15} SHMR sampled galaxies past $M_{\star} > 10^{11} M_{\odot}$ up into the BCG regime. This SHMR take the following form:

\begin{equation}\label{eq:star_halo}
     M_{\star}/M_h = 2 f_{1} \Bigg[\Big(\frac{M_{\star}}{M_1}\Big)^{-0.43}+\frac{M_{\star}}{M_1}\Bigg]^{-1} 
\end{equation}

Where $M_1$ is the transition or pivot halo mass, set to $10^{10.76} M_{\odot}$, and $f_{1}$ is the mass ratio at $M_1$, which is $f_1 = 0.0227$. Both SHMRs agree at $M_{\star} \sim 2.5\times 10^{10} M_{\odot}$, the ``knee" in the SHMR. Thus, for galaxies above this mass we use the \cite{Hudson15} SHMR, and for galaxies below this mass we use the \cite{Daneli23} SHMR, allowing for a smooth transition between the two SHMRs. We would like to note that, at this time, there has been no single study of the SHMR that has sampled galaxies from the full range of galactic stellar masses currently observed.

\subsection{Fits to The $M_{GCS}-M_h$ Relation}

\begin{center}
\begin{table*}[h!tb]
    \centering
    \caption{Linear Fit Solutions for $M_{GCS}-M_h$ Relation} \label{tab:ch3_fits}
    \begin{tabular}{ccc}
    \hline \hline
    Sample Combination & Slope & Intercept \\
    (1) & (2) & (3) \\
    \hline
    Dwarfs + Full Virgo + BCGs & $1.05 \pm 0.01$ & $-4.89 \pm 0.15$\\
    Non-UDG/ELSBG Dwarfs + Full Virgo & $0.96 \pm 0.02$ & $-4.01 \pm 0.20$ \\
    \hline
    Dwarf Catalog Only & $0.89\pm 0.03$ & $-3.21 \pm 0.28$\\
    Non-UDG/ELSBG Dwarfs Only & $0.92 \pm 0.02$ & $-3.53 \pm 0.26$\\
    \hline
    UDGs Only & $1.41 \pm 0.19$ & $-7.51 \pm 2.07$\\
    ESLBGs Only & $0.39 \pm 0.15$ & $1.89 \pm 1.45$\\
    UDGs + ESLBGs & $1.08 \pm 0.10$ & $-4.98 \pm 1.04$\\
    \hline
    \hline
    \end{tabular}
\item{} \footnotesize{\textit{Key to columns:} (1) Combination of observational samples used for the linear fits: either all of them, only one, or a combination of two; (2) The slope of the fit in log-log space; (3) The intercept of the fit in log-log space. }
\end{table*}
\end{center}

In Figure \ref{fig:h-MGC} a linear regression to the full combined sample for both galaxies with stellar masses below $M_{\star} < 2\times10^{10} M_{\odot}$ and for an additional sample up to the BCG regime. This relation from massive galaxies to dwarf galaxies was previously studied in \cite{Dornan25}, which used the full ACS Virgo Cluster Survey, a subset of Local Group dwarfs \citep{Harris13, Forbes18, Forbes20}, and a sample of BCGs. They found the $M_{GCS}-M_h$ relation for their full sample to behave as $\log(M_{GCS}) = (1.10 \pm 0.02) \log(M_h) - (5.64 \pm 0.89)$, with the BCG sample being offset above this fit due to the richer merger histories than the other galaxies in the sample.

We replaced the Local Group sample used in \cite{Dornan25} with our combined dwarf galaxy catalog and fit it alongside the same massive galaxies used by them. We find that the $M_{GCS}-M_h$ relation behaves as $\log(M_{GCS}) = (1.04 \pm 0.01) \log(M_h) - (4.75 \pm 0.15)$, resulting in a slightly shallower slope, but with intercepts in agreement within errorbars. When fitting for only the galaxies in our combined dwarf catalog, the $M_{GCS}-M_h$ relation behaves as $\log(M_{GCS}) = (0.89 \pm 0.02) \log(M_h) - (3.14 \pm 0.26)$, which is much shallower and offset lower than the global fit. This is driven by the exclusion of the positively offset BCG sample.

We would like to re-emphasize that these results are for the present-day $M_{GCS}$ estimates vs. the peak $M_h$ estimates, and the implications of this choice are discussed more thoroughly in Section \ref{subsec:past_present}. However, the behaviour of this $M_{GCS}-M_{h,peak}$ relation for our sample is actually quite similar to that of the $M_{GCS}-M_{h,z=0}$ relation found in \cite{Forbes18}. They found that, for the \cite{Georgiev10} sample dwarfs, when their halo masses were computed using the galaxies' HI gas kinematics, the behaviour of the $M_{GCS}-M_{h}$ relation was consistent with an extrapolation of the \cite{Spitler09} relation down to low halo masses. This resulted in an effective linear fit of $\log(M_{GCS}) = \log(M_h) - 4.15$, having roughly the same slope as the $M_{GCS}-M_{h,peak}$ relation found here, but with halo masses shifted lower by approximately 0.3 dex for the lowest mass dwarf galaxies. This subsample included only classical dwarfs, and thus the fit should be compared to that stated in line four of Table \ref{tab:ch3_fits}.

\subsection{Ultra Diffuse Galaxies and Extremely Low Surface Brightness Galaxies} \label{sec:UDGs/ELSBGs}

Within our combined dwarf galaxy catalog we also identify galaxies which meet the UDG classification criteria; with surface brightnesses above $\langle \mu_{g,0} \rangle > 24$ mag/arcsec$^2$ and effective radii above $R_e > 1.5$ kpc \citep{Dokkum15}. We also identify a subset of galaxies with effective radii below $R_e < 1.5$ kpc but with extreme surface brightnesses greater than $\langle \mu_{g,0} \rangle > 27$ mag/arcsec$^2$, which we will refer to as extremely low surface brightness galaxies (ELSBGs). Our aim is to investigate if these galaxies are significantly, systematically offset from the $M_{GCS}-M_h$ relation and if this is unique to UDGs or if it is a byproduct of the mechanisms that create significantly low surface brightness galaxies, regardless of size and concentration.

\begin{figure*}[h!tb]
    \centering
    \includegraphics[width=\textwidth]{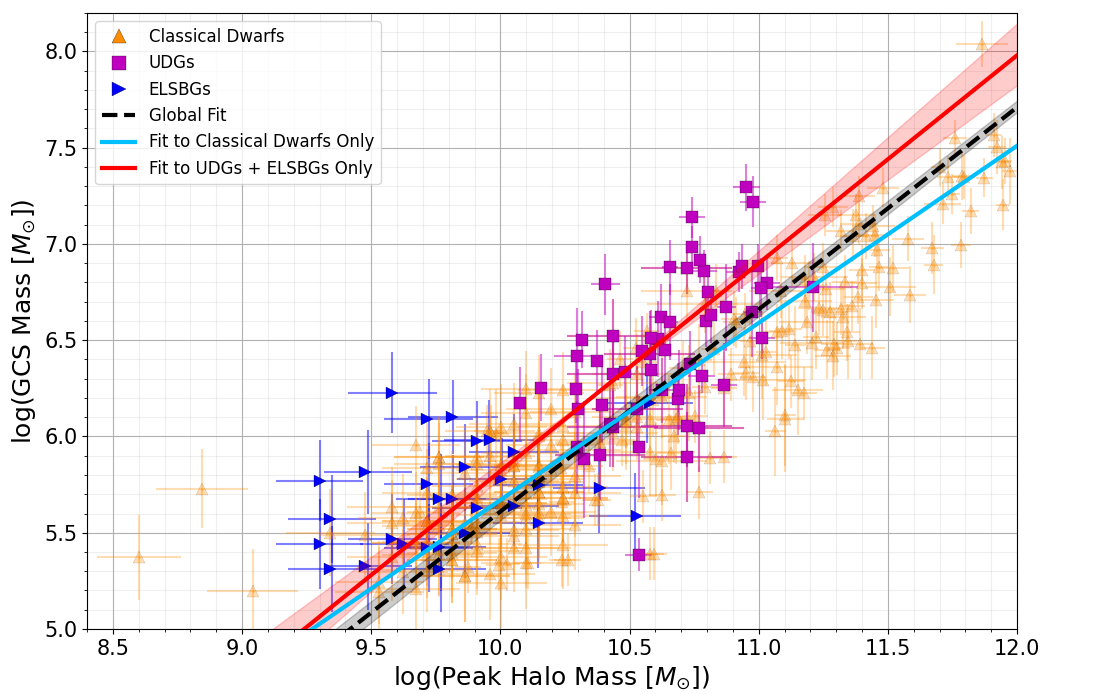}
    \caption{\label{fig:dwarf_fits} \footnotesize{$M_{GCS}$ vs peak $M_{h}$ for all galaxies in the combined sample. Orange triangles represent classical dwarfs, purple squares represent UDGs, and blue triangles represent ESLBGs ($\mu_{g,0}>27$). The black dashed line is the global $M_{GCS}$ vs peak $M_{h}$ relation fit, determined including massive galaxies. The blue line is the relation fit for only the dwarf galaxies in this combined catalog, and the red line is the relation fit for only UDGs and ELSBGs. The exact fits for these lines can be found in table \ref{tab:ch3_fits}}}
\end{figure*}

First, we note that galaxies classified as UDGs occupy a narrow range in peak halo masses but a comparatively large range in GCS masses. The UDGs in our sample have peak halo masses between $10^{10} M_{\odot} \leq M_{h,peak} \leq 10^{11} M_{\odot}$ and GCS masses between $2\times10^5 M_{\odot} \leq M_{GCS} \leq 2\times 10^7 M_{\odot}$. When determining the $M_{GCS}-M_h$ relation for our UDG subset we obtain a fit of $\log(M_{GCS}) = (1.42 \pm 0.18) \log(M_h) - (8.61 \pm 1.93)$. 

We also fit the UDG and ELSBG subsample together. The ESLBGs occupy a similar range in peak halo masses as the UDGs, although shifted lower, but have a much smaller range in GCS mass ($2\times 10^5 M_{\odot} \leq M_{GCS} \leq 2 \times 10^6 M_{\odot}$). When fitting the UDGs and the ELSBGs together we obtain a fit of $\log(M_{GCS}) = (1.02 \pm 0.08) \log(M_h) - (4.31 \pm 0.86)$, which has steeper slope compared to the non-UDG/ELSBG subsample, and shifted higher, shown in Figure \ref{fig:dwarf_fits}. 

This shows that many very low-mass galaxies in our sample have similar positions on the $M_{GCS}-M_h$ relation as the ELSBGs, while the UDGs systematically occupy higher GCS masses than classical dwarfs of the same peak halo mass. It should be noted, however, that the majority of galaxies in this combined catalog with peak halo masses below $M_{peak} < 10^{10} M_{\odot}$ are part of the Fornax Deep Survey and all have surface brightnesses above $\langle \mu_{g,0} \rangle > 24$ mag/arcsec$^2$ to begin with.

We find that low-mass ELSBGs lie in alignment with UDGs on the $M_{GCS}-M_h$ relation, with both being, on average, systematically positively offset from the relation in comparison to higher surface brightness dwarfs, in support of the result from \cite{Forbes+20}. We will note however, that while UDGs almost exclusively occupy the highest GC specific mass frequencies for their stellar mass range, other non-ELSBGs at very low stellar masses can also have as high GC specific frequencies as ELSBGs. We calculated the GC specific mass frequencies for all galaxies in our combined catalog, defined by equation \ref{eq:specific} below \citep{Zepf93,Peng08, Carlsten22}.

\begin{equation} \label{eq:specific}
    T_N = (10^9 M_{\odot})\times N_{GC}/M_\star
\end{equation}

\begin{figure*}[h!tb]
    \centering
    \includegraphics[width=\textwidth]{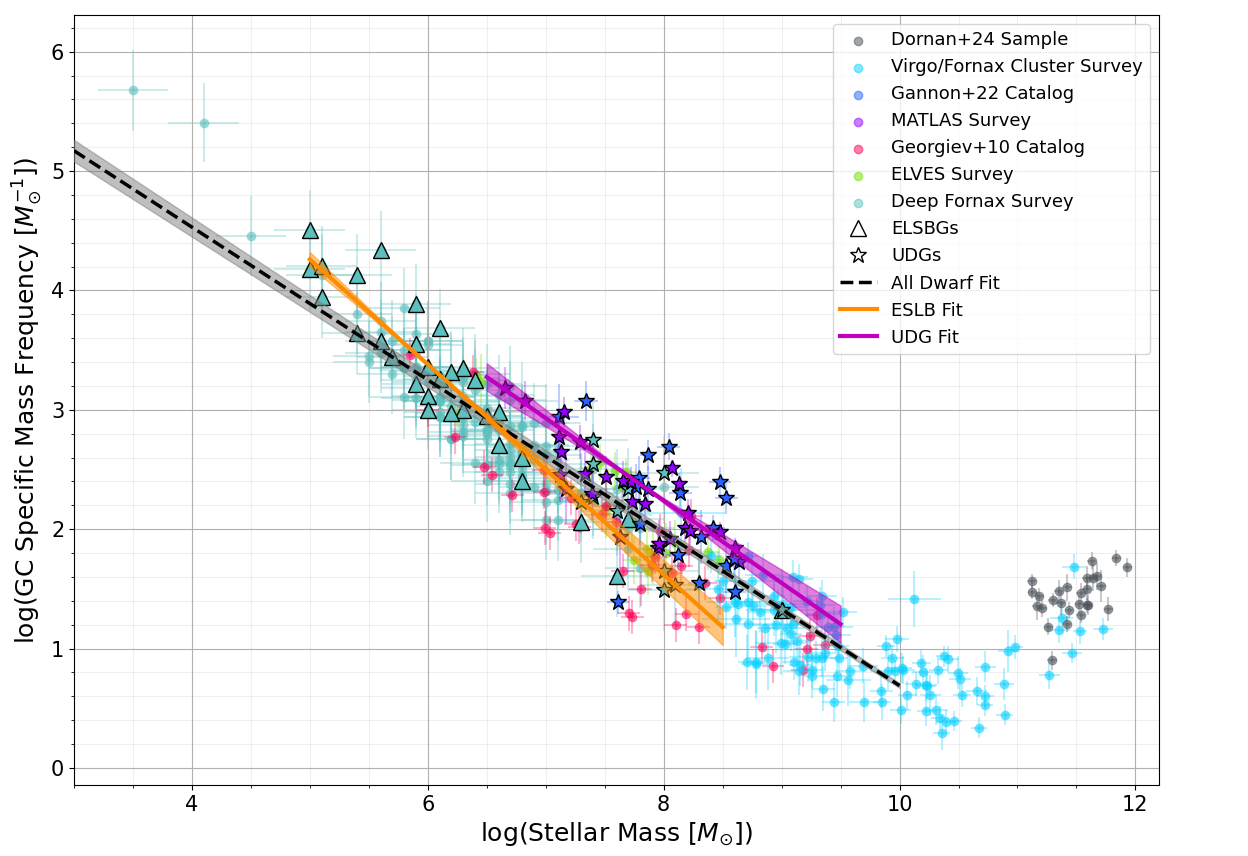}
    \caption{\label{fig:specific} \footnotesize{GC specifc mass frequency plotted against stellar mass for the combined dwarf galaxy catalog plus the massive Virgo cluster galaxies and the BCG sample. UDGs and ELSBGs are denoted by stars and triangles respectively. The fit for all dwarfs is plotted in black, the fit for UDGs only is plotted in purple, and the fit for ESLBGs only is plotted in orange.The exact fits for these lines can be found in table \ref{tab:specific}}}
\end{figure*}

\begin{center}
\begin{table*}[h!tb]
    \centering
    \caption{Linear Fit Solutions for $T_N-M_{\star}$ Relation} \label{tab:specific}
    \begin{tabular}{ccc}
    \hline \hline
    Sample & Slope & Intercept \\
    (1) & (2) & (3) \\
    \hline
    All Dwarfs & $-0.64 \pm 0.01$ & $7.09 \pm 0.12$\\
    ESLBGs Only & $-0.88 \pm 0.06$ & $8.66 \pm 0.36$ \\
    UDGs Only & $-0.69 \pm 0.09$ & $7.76 \pm 0.70$\\
    \hline
    \hline
    \end{tabular}
\item{} \footnotesize{\textit{Key to columns:} (1) Combination of sub-samples used for the linear fits; (2) The slope of the fit in log-log space; (3) The intercept of the fit in log-log space. }
\vspace{0.2cm}
\end{table*}
\end{center}

Figure \ref{fig:specific} plots GC specific mass frequency against stellar mass and illustrates this higher GC-richness for UDGs and ELSBGs, particularly the lowest mass ELSBGs, in our sample. We fit linear relations to all dwarfs in the catalog, the ESLBG sub-sample, and the UDG subsample in Figure \ref{fig:specific}, the parameters for which can be found in Table \ref{tab:specific}. We find that GC specific mass frequency scales tightly with stellar mass for the members of our dwarf galaxy catalog, with lower mass galaxies having higher $T_N$ values. This relation between $T_N$ and galaxy mass has been studied previously, sometimes comparing GC number specific frequency or GCS mass specific frequency, and sometimes comparing it to host galaxy absolute magnitude or dynamical mass. Regardless, a similar ``U-shape'' is consistently found (or in this case with the extension to such low masses, a ``hockey-stick shape''), with a negative linear slope for low-mass galaxies, followed by a an inflection around $M_{\star} \sim 10^{10} M_{\odot}$, and then a positive linear slope for high-mass galaxies \citep{Peng08, Harris13, Choksi19}. 

When studying the relation between $T_N$ and host galaxy stellar mass for low-mass galaxies, we notice that the UDG sub-sample has a similar slope to the dwarfs as a whole, but is shifted higher. This is consistent with UDGs, on average, having systematically higher GC specific mass frequencies than classical dwarfs. On the other hand, the ELSBG sub-sample has a steeper negative slope than the dwarfs as a whole, with only the lowest mass ESLBGs having $T_N$ values comparable to UDGs, and the highest mass ESLBGs having lower $T_N$ values than the average dwarf galaxy.

\section{Discussion} \label{sec:discussion}

In this section we will discuss the implications these results have on various fields of study for dwarf galaxy GCSs.

\subsection{Implications for UDGs and ELSBGs}\label{sec:udgs}

The properties of the GCSs of UDGs have been found to vary quite a bit, with many studies finding that UDGs can host extremely rich GC systems \citep{Mateu23, Gannon24, Janssens24, Forbes24}, or be GC-deficient for their masses \citep{Jones23, Buzzo25}. While this catalog does not have sufficient data to investigate the question of what may drive this difference in GC-richness for UDGs, others have found that this may be related to formation pathways. 

On average, UDGs with richer GC systems tend to be older, have lower metallicities, are found in denser environments like galaxy clusters, and have signs that their stellar populations formed early, quickly, and all at once throughout the galaxy \citep{Jones23,Mateu23, Buzzo25, Mateu25}. These could all be signs of a formation mechanism dependent on early tidal interactions or dwarf galaxy mergers which would ``puff-up" a classical dwarf galaxy through tidal heating and also trigger a burst of GC formation \citep{Mateu23,Fielder24}. 

While most UDGs found in galaxy clusters are associated with very early infall times, there exists no trend with GC-richness and time of infall \citep{Forbes23}. However, this was only investigated by categorizing UDGs as either GC-rich, with $N_{GC}>20$ or GC-poor, with $N_{GC}<20$. As can be seen in Figure \ref{fig:star-GC}, some UDGs with $N_{GC} \sim 10$ could still be considered GC-rich for their stellar mass. It would be useful for the analysis of \cite{Forbes23} to be re-conducted considering $N_{GC}/M_{\star}$ ratio, rather than simply $N_{GC}$.

Regardless, both GC-rich and GC-poor UDGs could still be formed via tidal heating and shocks due to cluster infall, but differ based on initial, internal galaxy properties that would encourage either GC formation and survival or GC destruction \citep{Forbes25}. Gas and GC density prior to infall could be a contributing factor, as higher gas densities would encourage GC formation \citep{Forbes25}, while higher densities in GC spatial distribution would encourage GC disruption from internal kinematics \citep{Hilario24}. 

This work indicates that similar mechanisms that produce GC-rich UDGs could potentially also produce the $N_{GC}/M_{\star}$ ratios and low surface-brightnesses seen in ELSBGs. This is similar to the results of \cite{Saifollahi25b}, who also found that diffuse and low surface-brightness galaxies in the Perseus Cluster had systematically higher $N_{GC}$ counts per unit stellar mass, similar to the UDGs studied in the same environment.

\subsection{Dwarf Galaxy GCLFs}

The GCLF is extremely important to consider in studies of dwarf galaxy GCSs. Not only is it used to estimate $N_{GC}$ through completness corrections, but its peak is also used to estimate average GC mass, and by extension total GCS mass for galaxies. However, due to the very low GC numbers hosted by dwarf galaxies, often it is not possible to construct an accurate GCLF for each individual galaxy. Instead, GCLFs are determined through stacking the GCS of a sample of dwarf galaxies.

This was the method applied by \cite{Miller07}, \cite{Jordan07}, and  \cite{Villegas10} to obtain the GCLFs used by the surveys in this catalog. However, these studies used samples of dwarfs to construct their GCLFs which did not go to the lowest dwarf galaxy luminosities included in this combined catalog. The lowest galaxy magnitude bin in \cite{Miller07} and \cite{Jordan07} was $M_V \sim -13$, while the lowest magnitude bin in \cite{Villegas10} was $M_V \sim -17.5$. More recently \cite{Saifollahi25} determined the stacked dwarf GCLF in the Fornax cluster using a dwarf galaxy sample ranging from $-17 \lesssim M_V \lesssim -14$. For context, the magnitude range for the galaxies in this catalog span from $-21.4 \leq M_V \leq -9.4$. 

Multiple studies have shown that both the peak magnitude and the dispersion in the GCLF can shift lower as a function of host galaxy luminosity \citep{Jordan07, Harris13}. As it currently stands, there is no broad study of the GCLF in galaxies with $M_V > -13$, of which nearly a third of this combined dwarf catalog would be classified. If the peak of the GCLF is actually at fainter magnitudes for these galaxies, it would mean that their $N_{GC}$ values could be under-estimations, or that their $M_{GCS}$ values could be over-estimations. 

In addition, all studies of the GCLF of dwarfs have been limited to the Fornax and Virgo galaxy clusters. This is due to the large number of dwarfs at similar distances, all of which have been included in deep, high resolution surveys with HST and Euclid. However, this can also introduce an environmental bias, with the GCLF of dwarf galaxies in the Local Volume currently poorly studied.

\subsection{A Present or Past Scaling Relation?} \label{subsec:past_present}

Finally, we would like to discuss the choice to have this scaling relation be dependent on peak halo mass rather than present halo mass. This choice immediately brings up the issue with comparing a past property of a galaxy (the highest halo mass hosted during its evolutionary history), against a present property (current GCS mass). 

The linear $M_{GCS}-M_h$ scaling relation likely originated immediately after GCS formation, and the scatter we observe today is due to galaxy evolution mechanisms that have affected both dark matter halo and GCS masses \citep{Choksi19}. Ideally, we would also be able to plot ``original" GCS mass alongside peak halo mass, but that is not currently possible observationally. However, the evolution of the GCLF has been studied through simulations, and has found that the brightest, most massive half of the GCLF does not change significantly, while the low-mass GC population gets dynamically reduced \citep{Li14,Marta22}. The result is that the total GCS mass, dominated by the mostly undisturbed high-mass GCs, changes slowly since the time at which the host galaxy reached its peak halo. Through limiting the halo mass to what it would have been originally, we can better study the variations in GCS mass per unit halo mass that have since occurred in dwarf galaxies.

Several different theoretical studies of the $M_{GCS}-M_h$ relation have found that for their fiducial models, when GC formation is limited to above a high gas surface density threshold, that a declining $M_{GCS}/M_h$ ratio for dwarf galaxies is predicted, with the relation for dwarfs being steeper, and the galaxies themselves sitting lower than is found in Figure \ref{fig:dwarf_fits} \citep{El-badry19, Choksi19, Valenzuela21}. However, allowing for more varied GC formation \citep{El-badry19}, or accounting for GC formation via gas-rich galaxy mergers \citep{Valenzuela21} re-creates the larger spread in the $M_{GCS}-M_h$ relation for dwarfs that is observed, and recovers the continuation of linearity in the $M_{GCS}-M_h$ relation as well. While galaxy mergers have been shown to affect the scatter in this relation, less has been done to study the effect of tidal disruption. 

By studying the $M_{GCS}-M_{peak}$ relation we can begin to form an observational understanding of the properties and evolutionary histories of dwarf galaxies which relate to the growth or destruction of the GCSs over time. This is turn can inform future studies of GCS evolution as a product of galaxy-galaxy interactions.

\section{Summary} \label{sec:summary}

In this study we compiled a literature catalog of six systematic surveys of dwarf galaxy GCSs as well as a previous literature catalog encompassing various individual galaxy studies. For this combined catalog we determined both GCS masses and peak halo masses for each galaxy included using a consistent and standardized method to ensure that comparisons of these masses between surveys would be accurate. The result is the most complete, standardized study of the $M_{GCS}-M_h$ relation for the entire dwarf galaxy regime to date. We summarize the results of our analysis of this catalog below:

\begin{enumerate}
    \item We find that for galaxies with stellar masses $M_{\star} \lesssim 5 \times 10^{9} M_{\odot}$ the linearity in the $M_{GCS}-M_h$ relation holds, with the majority of the increasing scatter compared to higher mass galaxies being driven by UDGs and ELSBGs.
    \item When excluding BCGs, UDGs, and ELSBGs, which we find to have systematically higher GC counts per unit mass, the slope of the $M_{GCS}-M_h$ relation is consistent from dwarf galaxies to massive galaxies up to $M_h \sim 10^{13}M_{\odot}$, taking on the form: $\log(M_{GCS}) = 0.95 \log(M_h) - 3.85$.
    \item We found that, for dwarf galaxies, GC specific mass frequency ($T_N$) scales tightly with host galaxy stellar mass, with lower mass dwarfs having consistently higher $T_N$ values.
    \item We found that UDGs follow the same $T_N-M_{\star}$ relation as for classical dwarfs, but shifted to systematically higher $T_N$ values. ESLBGs, on the other hand, follow a steeper $T_N-M_{\star}$ relation compared to classical dwarfs, with higher mass ESLBGs having lower $T_N$ values than the average classical dwarf of the same stellar mass.
    \item While the UDGs in our sample occupied a very wide range in GCS masses, galaxies with the highest $T_N$ values for their stellar masses were exclusively UDGs, not classical dwarfs. In addition, the lowest mass ELSBGs had similar $T_N$ values to UDGs, although having smaller effective radii. This implies that similar mechanisms that form GC-rich UDGs could also form low-mass ELSBGs.
\end{enumerate}

\section{Future Work} \label{sec:future}

There are further steps required to expand and improve upon this dwarf galaxy GCS catalog in order to draw wider conclusions about what dictates where dwarf galaxies lie on the $M_{GCS}-M_h$ relation. Importantly, galaxies which host no GCs should be included to gain a more complete picture of the scatter observed in the scaling relation. This will also allow for the analysis of what galaxy properties are most closely correlated with a lack of GCs, as opposed to the GC-hosting dwarfs in this catalog currently.

In addition, more consistent data must be collected for global dwarf galaxy properties which could be indicators of the galaxies' GCS evolution histories, or of the galaxy's hospitality to GC formation and survival. Properties of note would be gas mass, gas surface density, and time of infall within their galaxy clusters. Systematic surveys, such as Euclid and LSST, could provide some of this data in the near future.

\section{Acknowledgments}

We would like to thank Michelle Collins, Justin Read, and Jonnah Gannon for their helpful conversations and insights that aided this work. We would also like to thank the anonymous reviewer, as well as Alison Sills, Laura Parker, and Rupali Chandar for their excellent feedback and comments.

This work was supported by a Discovery Grant to WEH from the Natural Sciences and Engineering Research Council of Canada (NSERC). 

\appendix

\begin{turnpage}
\begin{table*}[h!tb]\label{app:Gannon_papers} 
    \footnotesize
    \centering
    \caption{Gannon Catalog Citations}
    \begin{tabular}{ll}
    \hline \hline
    Name & Citation \\
    \hline
    DF44 & \citeauthor{vanDokkum2016}(\citeyear{vanDokkum2016},\citeyear{vanDokkum2017}, \citeyear{vanDokkum2019b}), \cite{ Gannon2021, Villaume2022, Webb2022, Saifollahi2022} \\
    DF07 & \cite{vanDokkum2015, Gu2018, Lim2018, Saifollahi2022, FerreMateu2023}  \\
    DF17 & \cite{Peng2016, Beasley2016, vanDokkum2017, Gu2018, Saifollahi2022}  \\
    DFX1 & \cite{vanDokkum2017, Gannon2021, Saifollahi2022, FerreMateu2023}  \\
    DGSAT-I & \cite{MartinezDelgado2016,MartinNavarro2019, Janssens2022}  \\
    Hydra-I UDG 11 & \cite{Iodice2020,Iodice2023} \\
    NGC 1052-DF2 & \cite{vanDokkum2018,Fensch2019,Danieli2019, Shen2021,Shen2023}  \\
    NGC 5846 UDG1 & \cite{Forbes2019}, \citeauthor{Muller2020} (\citeyear{Muller2020} \citeyear{Muller2021}), \cite{,Forbes2021, Danieli2022,FerreMateu2023} \\
    NGVSUDG-19 & \cite{Lim2020,Toloba2023}  \\
    NGVSUDG-20 & \cite{Lim2020,Toloba2023}  \\
    Sag dSph & \cite{mcconnachie2012,Karachentsev2017,Forbes18}  \\
    VCC 1017 & \cite{Lim2020,Toloba2023}  \\
    VCC 1052 & \cite{Lim2020,Toloba2023}  \\
    VCC 1287 & \cite{Beasley2016}, \citeauthor{Gannon2020} (\citeyear{Gannon2020}, \citeyear{Gannon2021}), \cite{Lim2020,Toloba2023}  \\
    VCC 615 & \cite{Lim2020,Toloba2023} \\
    VCC 811 & \cite{Lim2020,Toloba2023}  \\
    VLSB-B & \cite{Toloba2018,Lim2020,Toloba2023}  \\
    VLSB-D & \cite{Toloba2018,Lim2020,Toloba2023}  \\
    WLM & \cite{mcconnachie2012,Forbes18}  \\
    Y358 & \cite{vanDokkum2017,Lim2018,Gannon2023}  \\
    \end{tabular}
\end{table*}
\end{turnpage}

\bibliography{paper}{}
\bibliographystyle{aasjournal}

\end{document}